CRYSTALLIZATION
# Colloidal suspense

According to classical nucleation theory, a crystal grows from a small nucleus that already bears the symmetry of its end phase — but experiments with colloids now reveal that, from an amorphous precursor, crystallites with different structures can develop.

### László Gránásy & Gyula I. Tóth

Colloidal suspensions consist of small, typically micrometre-sized, particles floating in a carrier fluid: their ability to crystallize is a feature often used to explore the process of nucleation, during which tiny crystallites form through thermal fluctuations of the non-equilibrium (undercooled or supersaturated) fluid. In a study published in *Nature Physics*, Peng Tan and colleagues[1] uncover the possible pathways through which a colloidal suspension can evolve from the liquid to the solid state. Although mounting evidence in recent years has supported a process of two-stage nucleation in simple liquids[2-8], the new work provides long-awaited details on its structural aspects.

In the colloidal crystallization experiments performed by Tan and colleagues[1], the solvent is a mixture of non-polar and weakly polar solvents. By changing the polar/non-polar ratio of the solvent's components, the (soft-repulsive) interaction between the colloidal particles can be tuned. Essentially, this means that the crystal structure of the nucleating state can be regulated: body-centred cubic (bcc) or face-centred cubic (fcc). The authors have also investigated crystallization in a hard-repulsive system, where the end-state is a random mixture of fcc and hexagonal close-packed (hcp) symmetries — known as random hcp. Using three-dimensional confocal microscopy, they analysed the evolving configurations in terms of so-called bond-order parameters — rotationally invariant measures for the structure of the first- and second-neighbour shells around any given particle[9]. From an examination of bond-order parameters and solid-bond numbers (for a given particle, the solid-bond number is the number of neighbours lying within the crystalline phase), Tan *et al.* distinguished three different states of crystallizing matter: liquid, precursor and nucleus. To compute the bond-order parameters, they introduced a method that corrects for the randomizing effect of thermal fluctuations, yielding particles with temporary, non-crystalline coordination even in crystalline or crystal-like regions — an approach that helps to identify the local structure.

From these experiments, it is clear that in all these colloid systems the observed crystalline phases nucleate through an intermediate step. First, an amorphous precursor forms with mainly hcp-like short-range order, but also with some minor bcc- and fcc-like content (Fig. 1). Then, during the actual crystallization process, the precursor develops into a crystalline phase that is determined by the nature of the interparticle interaction (soft- or hard-repulsive). Fcc-, bcc- or hcp-like precursors can evolve into crystalline nuclei of any of these symmetry types. The densities of the precursors lie between those of the crystalline and the liquid phases. However, the spatial distributions of density and structure in the crystallizing liquid do not seem to correlate — a finding that deserves further investigation. And although these results extend our knowledge of crystal nucleation substantially, they also raise intriguing questions: why does one have non-crystalline precursors in the first place, and how general is the presence of amorphous precursors in crystal nucleation?

Let us address the second question first. There are several cases in which a structurally disordered precursor has been seen to assist the formation of a crystalline phase: optical studies on colloids, which show that crystal nucleation happens inside dense, low-symmetry amorphous precursors[3-5]; Monte Carlo simulations of the hard-sphere system, in which dense, amorphous clusters form first, assisting the nucleation of crystallites[6]; theoretical studies that predict a dense fluid precursor for globular proteins and the Lennard–Jones system[7]; and dynamical density functional studies based on the so-called phase-field crystal (PFC) technique (a molecular theory of crystalline freezing with an effective potential approximating the interactions between the weakly charged colloids), which demonstrate that crystallization starts with the nucleation of amorphous droplets of density between that of the bulk crystalline and liquid phases[8].

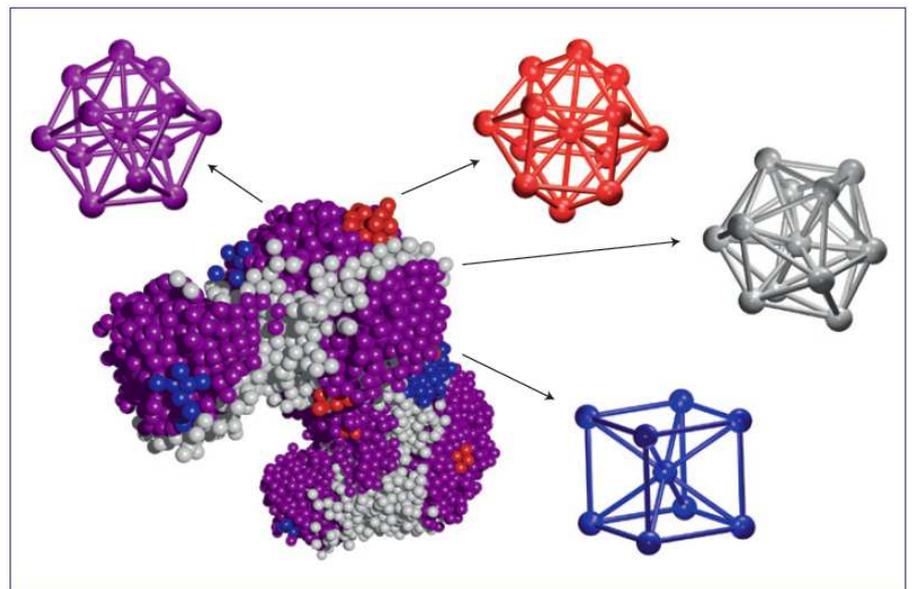

**Fig. 1** | Crystal nucleation in colloidal suspensions is a two-step process, the first of which involves the formation of amorphous precursors with different types of short-range order. The existence of domains with hcp-like (purple), fcc-like (red) and bcc-like (blue) order is confirmed by Tan *et al.*[1], who have analysed sequences of microscopy images taken from micrometre-sized spherical colloids (represented as coloured balls here) in a solvent. The presence of a domain with liquid-like short-range order (grey) may be expected on the basis of dynamical density functional theory[8]. In the second step, the precursors evolve into nuclei with fcc, bcc or hcp structure. The enlarged purple, red and blue clusters show a central particle surrounded by its nearest neighbours in ideal hcp, fcc and bcc environments, respectively. The crystal-like short-range order can be obtained by distorting these ideal crystalline first-neighbour clusters. To exemplify the liquid-like neighbourhoods, a cluster with neighbouring particles lying on the vertices of a distorted icosahedron is shown.

Bond-order analysis of PFC results reveals hcp-like short-range ordering for the precursor; the corresponding order-parameter map is less structured than for colloid experiments, however. In addition, in the PFC simulations a substantial amount of another type of precursor forms, having 'liquid-like' short-range order: its bond-order distributions coincide with those of the bulk liquid. Accordingly, this second type of amorphous precursor cannot be distinguished from the liquid by bond-order analysis. So, what Tan et al.[1] identified as liquid could in part consist of solid precursors of liquid-like short-range order, and, in the present state of affairs, the amorphous precursors initiating crystallization could still be mixtures of domains of liquid- and crystal-like neighbourhoods.

But why should such amorphous precursors form anyway? A general argument has been given by Lutsko and Nicolis[7], suggesting that it is easier, from a free-energy-minimization point of view, to crystallize by passing through a metastable dense fluid state rather than ordering and densifying simultaneously as assumed in the classical picture of crystallization. A recent theoretical tour de force using the density-functional approach (the only practicable molecular theory of crystalline freezing) offers another clue[10]: in highly non-equilibrium liquids, solidification starts with spherical density waves — resulting in a kind of 'onion structure' — as they are energetically favourable to small crystalline clusters. This finding may offer a natural explanation of the amorphous precursor: except for the first-neighbour clusters, these concentric density waves are incompatible with a long-range crystalline structure. The interference of the onion structure with thermal density fluctuations may account for the observed preference for early-stage amorphous solidification.

Putting everything together, one has the following scenario in the language of continuum theory (in terms of time-averaged particle density). First, the highly symmetric homogeneous liquid state (characterized by a uniform particle density) loses its translational symmetry when the onion structures form. Then, all symmetries disappear when the amorphous precursor forms (which, however, may already contain crystal-like regions), and eventually the highly symmetric crystalline phase develops. So symmetry breaking during the crystallization of a liquid seems to be a complex process, in which the transition between two highly symmetric states happens via intermediate states of lower symmetry. Further experimental, numerical and theoretical work is still needed to merge these pieces of information into a unified picture of nucleation, in which structural analyses of the type performed by Tan et al.[1] will certainly have their place.

*László Gránásy[1,2] and Gyula I. Tóth[1] are at [1]Computational Materials Science & Physics Group, Wigner Research Centre for Physics of the Hungarian Academy of Sciences, H-1525 Budapest, Hungary, [2]Brunel Centre for Advanced Solidification Technology, Brunel University, Uxbridge UB8 3PH, UK. e-mails: granasy.laszlo@wigner.mta.hu; toth.gyula@wigner.mta.hu*

References
1. Tan, P., Xu, N. & Xu, L. *Nature Phys.* **10,** 73–79 (2014).
2. Kawasaki, T. & Tanaka, H. *Proc. Natl Acad. Sci. USA* **107,** 14036–14041 (2010).
3. Schöpe, H. J., Bryant, G. & van Megen, W. *Phys. Rev. Lett.* **96,** 175701 (2006).
4. Zhang, T. H. & Liu, X. Y. *J. Am. Chem. Soc.* **129,** 13520–13526 (2007).
5. Savage, J. R. & Dinsmore, A. D. *Phys. Rev. Lett.* **102,** 198302 (2009).
6. Schilling, T., Schöpe, H. J., Oettel, M., Opletal, G. & Snook, I. *Phys. Rev. Lett.* **105,** 025701 (2010).
7. Lutsko, J. F. & Nicolis, G. *Phys. Rev. Lett.* **96,** 046102 (2006).
8. Tóth, G. I., Pusztai, T., Tegze, G., Tóth, G. & Gránásy, L. *Phys. Rev. Lett.* **107,** 175702 (2011).
9. Lechner, W. & Dellago, C. *J. Chem. Phys.* **129,** 114707 (2008).
10. Barros, K. & Klein, W. *J. Chem. Phys.* **139,** 174505 (2013).